\newcommand{\insertps}[2]{
  \begingroup                   
     \def\epsfsize##1##2{#1##1} 
     \epsfbox{#2}               
  \endgroup}
\newcommand{\SY}{\scriptstyle}

\documentstyle[preprint,tighten,floats,pre,aps,epsf]{revtex}

\begin{document}
\title{Deformation of polymer films by bending forces}
\author{Gerald P\"atzold\thanks{Contact:
  {\tt paetzold@tphys.uni-heidelberg.de},
  {\tt http://wwwcp.tphys.uni-heidelberg.de/}},
  Thorsten Hapke, Andreas Linke, and Dieter W.~Heermann}
\address{Institut f\"ur Theoretische Physik, Universit\"at Heidelberg,
  Philosophenweg 19, D--69120 Heidelberg, Germany
  and Interdisziplin\"ares Zentrum f\"ur wissenschaftliches Rechnen
  der Universit\"at Heidelberg}
\date{\today}
\maketitle
\begin{abstract}
We study the deformation of nano--scale polymer films which are subject
to external bending forces by means of computer simulation.
The polymer is represented by a generalized bead--spring--model,
intended to reproduce characteristic features of n--alkanes. The film is
loaded by the action of a prismatic blade which is pressed into the polymer
bulk from above and a pair of columns which support the film from below.
The interaction between blade and support columns and the polymer is
modelled by the repulsive part of a Lennard-Jones potential. For different
system sizes as well as for different chainlengths, this nano--scale
experiment is simulated by molecular dynamics methods. Our results allow
us to give a first characterization of deformed states for such films. We
resolve the kinetic and the dynamic stage of the deformation process in
time and access the length scale between discrete particle and continuum
mechanics behaviour. For the chainlengths considered here, we find that
the deformation process is dominated by shear. We observe strangling
effects for the film and deformation fluctuations in the steady state.
\end{abstract}

\section*{The film bending experiment}

\begin{figure}
  \begin{center}
    \begin{minipage}{\textwidth}
      \insertps{1.00}{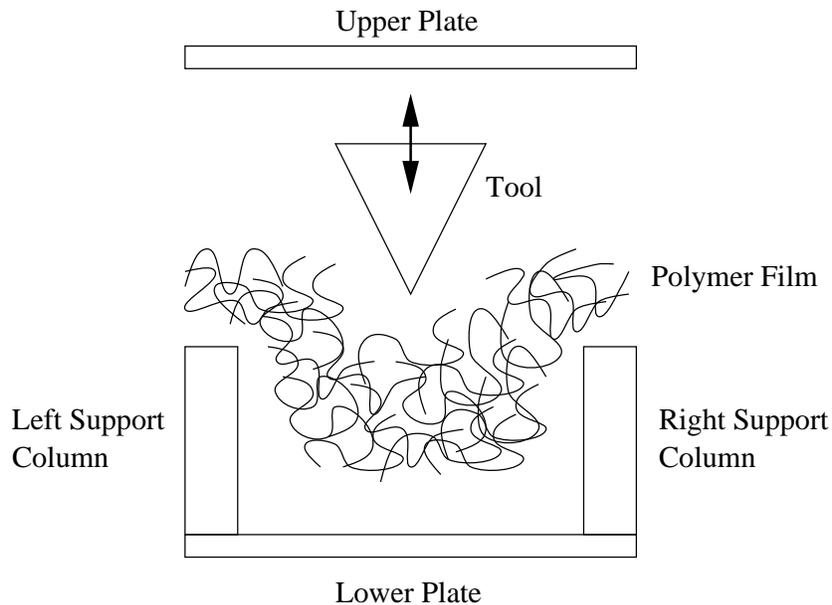}
    \end{minipage}
  \end{center}
  \caption{The building blocks of the computer experiment.}
  \label{fig-01}
\end{figure}

Nano--scale deformation processes are difficult to deal with. Typical
systems may consist of $10^4$ to $10^6$ interacting particles but are
still too coarse to apply continuum mechanics concepts directly. Computer
simulation can fill the gap. In the following, we consider the effect of
bending forces on  polyethylene films with an equilibrium thickness
of approximately 75 {\AA}. Our perspective is somewhat different from that of
a large group of computer studies on polymers which mainly address the bulk
properties of the material. We do not restrict ourselves to equilibrium
situations or perturbations around  equilibrium, and it is not clear from
the beginning if our results can be interpreted in terms of a linear
viscoelastic response. As can be seen in Fig.~\ref{fig-01},
our simulations take place in a more complicated
geometry than usual and the material undergoes considerable deformations.
We consider surface effects and the interaction between polymer and rigid
bodies which are pressed into the bulk. The problem as a whole also involves
the search for a set of parameters which qualify for a concise description
of what we are looking at.

The present line of work started with the indentation simulations of
Hapke \cite{Hap95a}. In these computer experiments, a rigid tip is
pressed into the material to probe the resistance of the polymer surface.
The indentation depth of the tip and a number of related parameters are
recorded and analyzed. It became clear that this approach to nano--mechanics
can be extended to more complicated situations. In addition to the
investigation of nano--indentation and its connection to surface hardness
and degradation, we can also contribute to fields of interest in
nano--technology, like nano--bending, nano--forging, and nano--milling.
In the next sections, we first describe the modelling of the bulk polymer
and the interaction between polymer and solid bodies. We sketch the
geometry of the computer experiment and mention some essential features
of the simulation program. We then indicate a simple way to analyze the
data from the computational deformation experiments and apply the method
to data from various simulation runs. In particular, we look at the cross
section displacement, the variable film thickness, the time development
of the indentation depth, and the local density field in the bulk next
to the invading body. Finally we summarize our findings and outline some
possibilities for further research.

\subsection*{The polyethylene model}

In computer experiments on the nano--mechanics of amorphous media,
one typically has to model a number of different subsystems, including
bulk polymer (chain molecules or polymer networks), free polymer surfaces,
the interaction between polymer and tools (possibly manufactured from
metal), and  the behaviour of the tools themselves \cite{Hap95a}. Each
subsystem can be modelled on a different level of sophistication. We have
to follow the motion of the monomers in the polymer chains, but are not
interested in the dynamics of the electron gas in the metal tool.
In what follows, we employ a standard model for chain molecules.
New technical aspects of our work mainly concern the implementation of
rigid geometries representing the solid tools and supports which can act
onto the material. For the polymer, however, we use a common bead--spring
model with some extensions to capture the essential features of
polyethylene chains \cite{Hap95a}. In addition to harmonic chain forces
which keep the bond lengths next to the equilibrium value, we model the
fluctuation of bond angles, again   by a quadratic potential. Between
monomers which do not participate in mutual bond length or bond angle
interactions, Lennard--Jones forces are acting, both to model an excluded
volume effect and to hold the polymer film together (the total energy in
all our simulations is negative, the system is in a bound state). Note that
we neglect any torsional potential in the present study.
To be explicit, the Hamiltonian of the polyethylene model is of the form
\begin{eqnarray}
  H &=& H_{\rm \SY bondlength} + H_{\rm \SY bondangle} + H_{\rm \SY LJ}
    \, , \nonumber \\
  H_{\rm \SY bondlength} &=& \sum_{\rm \SY bonds} \frac{k_b}{2}
    \left( l_{\rm \SY bond}
    - l_0 \right)^2 \, , \nonumber \\
  H_{\rm \SY bondangle} &=& \sum_{\rm \SY angles} \frac{k_\theta}{2} \left( 
    \cos \theta_{\rm \SY angle} - \cos \theta_0\right)^2 \, , \\
  H_{\rm \SY LJ} &=& \sum_{\rm \SY pairs \; of \atop \rm \SY monomers}
    4 \epsilon \left[ \left( \frac{\sigma}{r_{\rm \SY pair}} \right)^{12} -
    \left( \frac{\sigma}{r_{\rm \SY pair}} \right)^{6} \right] \, \nonumber .
\end{eqnarray}
Models of this kind have been described at various places in the
literature (e.g.~Ref.~\cite{All87a}, and, for specific details,
Ref.~\cite{Hap95a}), so we can be short here. Let us only mention
that we implement the Lennard--Jones interaction using the
linked cell algorithm \cite{All87a}. At the cutoff distance,
we restore continuity
in energy and force by appropriate shifts. The time stepping is done
through the velocity form of the Verlet algorithm.
The simulation program can be run in various major modes: pure molecular
dynamics, molecular dynamics with velocity scaling (as a brute force
approach to guarantee a constant temperature), Hybrid Monte Carlo
(a combination of Monte Carlo and molecular dynamics,
where the molecular dynamics part acts as the event generator), and
Brownian dynamics (additive Langevin forces simulate the coupling to a
heat bath). Also a Hoover thermostat has been implemented. The simulations
reported on below were performed using pure molecular dynamics with
the Hoover thermostat active. A number of parameters relevant to all these
runs is given in Tab.~\ref{table-01}.
\begin{table}
  \caption{Parameters of the polyethylene model.}
  \label{table-01}
  \begin{tabular}{lll}
    Lennard--Jones energy, $\epsilon$ & $8.3027 \cdot 10^{-22}$ J
      & 5.18 meV \\
    Lennard--Jones length, $\sigma$ & 380 pm
      & 3.8 {\AA} \\
    monomer mass (${\rm CH}_2$ group), $m$ & $2.3248 \cdot 10^{-26}$ kg
      & 14 atomic units \\
  \tableline
    unit of temperature, $\epsilon / k_{\rm B}$ & 60.1357 K \\
    unit of mass density, $m / \sigma^3$ & 423.6687 ${\rm kg}/{\rm m}^3$
      & 0.4237 ${\rm g}/{\rm cm}^3$ \\
    unit of time, $\left(m \sigma^2 / \epsilon \right)^{1/2}$ & 2.0108 ps \\
    unit of velocity, $\left( \epsilon / m \right)^{1/2}$ & 188.9822 m/s \\
    unit of force, $\epsilon / \sigma$ & 2.1849 pN
      & 1.36 ${\rm meV} / {\rm {\AA}}$ \\
    unit of spring constant, $\epsilon/\sigma^2$ & $5.7498 \cdot 10^{-3}$ N/m
      & 0.36 ${\rm meV} / {\rm {\AA}}^2$ \\
    unit of pressure, $\epsilon / \sigma^3$ & 151.3103 bar
      & 0.09 ${\rm meV} / {\rm {\AA}}^3$ \\
  \tableline
    temperature, $T$ & 361 K \\
    bond length, $l_0$ & 152 pm
      & 1.52 {\AA} \\
    spring constant (bond length), $k_b$ & $5.7498 \cdot 10^{1}$ N/m
      & 3.59 ${\rm eV} / {\rm {\AA}}^2$ \\
    bending constant (bond angle), $k_\theta$ & $8.3027 \cdot 10^{-19}$ J
      & 5.18 eV \\
    simulation time step, $\Delta t$ & 2.0108 fs \\
    external force on tool, $F$ & 0.25 nN
      & 0.16 ${\rm eV} / {\rm {\AA}}$ \\
    box depth (y)                            & 6.1 nm & 61 {\AA} \\
    box heigth (z) (accessible, not filled)  & 15.2 nm & 152 {\AA} \\
    equilibrium film thickness               & $\approx \text{7.5 nm}$
      & $\approx \text{75 {\AA}}$ 
  \end{tabular}
\end{table}

\subsection*{Interaction between polymer bulk and solid bodies}

We model the interaction between
polymer and solid bodies (tools and supports) solely by the repulsive
($r^{-12}$) part of the Lennard--Jones potential. Moreover, we use the
same parameters (energy scale $\epsilon$ and length scale $\sigma$) as
in the polymer bulk. At first glance, such an approach might not look
appropriate at all and it certainly needs some justification.
Our main argument is that we are just interested in what is independent
of specific interaction models. Of course this presupposes that there are
aspects of nano--deformation which do not depend on  details like the exact
potentials between organic molecules and metal surfaces. Work in the latter
direction has been carried out by others \cite{Cha90a}, and it is certainly
the next step to utilize their results. However, we are here concerned with
issues which we think are both fundamental to the problem of nano--deformation
and merely geometrical in nature. We want to observe what happens on
a nano--scale if a material like polymer is displaced through the invasion of
a compact body with a given shape.
Within this approach, we cannot expect to resolve any interaction zones
between polymer and tool. But what we miss are effects in a boundary layer
on an even smaller scale which are not suspicious to dominate the overall
behaviour.  One should note, however, that we are not able to model any
adhesion effects which become important if we pull the tool out of the
polymer. A straightforward way to capture those aspects is to include
the attractive part of the Lennard--Jones potential into the modelling.

In our computer experiments, we meet two types of rigid geometries
which interact with the polymer film: fixed and movable ones. Fixed are
the bottom and top plates and the two columns which support the film
(see Fig.~\ref{fig-01}). They repell monomers but do not feel the reaction
forces. The tool, a prismatic blade, belongs to the second category and
is movable. It is modelled as a massive body of $10^4$ monomer masses which
obeys Newtonian dynamics. The motion of the tool can therefore be
naturally included into the overall time stepping algorithm for second
order dynamics (here the velocity form of the Verlet algorithm).
As a program option, the motion of the tool can also be
controlled by distance \cite{Hap95a}. This allows us to drive the
tool to a certain position in the bulk and to probe
the reaction forces, but we did not use this feature here.
A number of tool parameters can be varied, including the tip radius
(here we use an acute tip with a formal tip radius of zero),
the tip opening angle (here 30 degrees), the insertion force (here 0.25 nN),
the tip mass (here $1.4 \cdot 10^5$ atomic units = $2.3248 \cdot 10^{-22}$ kg),
and the tool's angle of attack (here straight from above).
In any case, whether by prescribed motion or under the action of an
external force, pressing the tool into the bulk adds energy to the
polymer subsystem. The use of some kind of thermostat is therefore
mandatory to achieve an isothermal simulation for the polymer.
In Fig.~\ref{fig-01}, a cross section through the experiment in
the x/z--plane is shown. As mentioned, the boundary in the height (z)
direction consists of repelling plates. In the lateral directions,
refered to as width (x) and depth (y), we use common periodic boundary
conditions. The depth is a neutral direction according to this
setup and is exploited for additional averaging.

\section*{Results from the computer experiments}

\begin{figure}
  \begin{center}
    \begin{minipage}{\textwidth}
      \insertps{0.7}{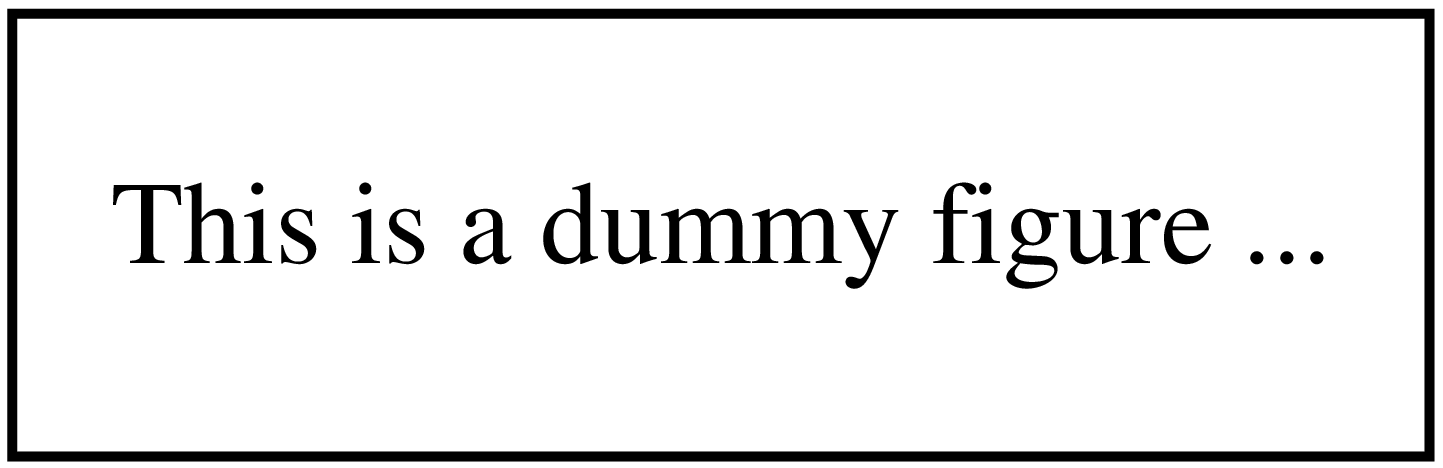}
    \end{minipage}
  \end{center}
  \caption{A small system of width 60 {\AA} with 8220 chains of length 60,
    shown 1.6 ns after the start of the bending experiment.}
  \label{fig-02}
\end{figure}
%
\begin{figure}
  \begin{center}
    \begin{minipage}{\textwidth}
      \insertps{0.7}{dummy.eps}
    \end{minipage}
  \end{center}
  \caption{A medium system of width 120 {\AA} with 16400 chains of length 20,
    shown 0.8 ns after the start of the bending experiment.}
  \label{fig-03}
\end{figure}
We report on computational bending experiments for systems of three
different sizes and refer to them as to the small (Fig.~\ref{fig-02}),
medium (Fig.~\ref{fig-03}), and large configurations.
Simulation parameters which apply to all systems have
been summarized in Tab.~\ref{table-01}. The systems differ in their
width and the number of monomers considered. For each size, results
for chains built from 20 and 40 monomers (small system also
60 monomers) are available. Note that for each configuration size,
we keep the number of monomers constant, so going to longer chains
means a reduction in the chain number. More details are given in
Tab.~\ref{table-02}.
%
\begin{table}
  \caption{The various simulated polyethylene systems.}
  \label{table-02}
  \begin{tabular}{ll}
    Small polyethylene system.\\
  \tableline
    number of chains (chainlengths 20, 40, and 60) & 410, 205, and 137 \\
    number of monomers                       & 8200 \\
    box width (x)                            & 61 {\AA} \\
    total simulation time                    & $\approx \text{4 ns}$ \\
  \tableline
    Medium polyethylene system.\\
  \tableline
    number of chains (chainlengths 20 and 40) & 820 and 410 \\
    number of monomers                        & 16400 \\
    box width (x)                             & 122 {\AA} \\
    total simulation time                     & $\approx \text{1 ns}$ \\
  \tableline
    Large polyethylene system.\\
  \tableline
    number of chains (chainlengths 20 and 40) & 1640 and 820 \\
    number of monomers                        & 32800 \\
    box width (x)                             & 243 {\AA} \\
    total simulation time                     & $\approx \text{? ns}$
  \end{tabular}
\end{table}

\subsection*{The characterization of deformation}

Although the blade is driven by a force of only 0.25 nN, the polymer
film undergoes considerable deformations, especially in the medium and
large systems. In continuum mechanics, the standard method to describe any
deformation is to use three--dimensional vector fields. In order to
proceed that way, one has to define so--called material points which
can be labeled and followed in space during the deformation history.
One might think about them as marked pieces of matter. The vector field
then encodes the difference in position of the material points between
some reference and the current configuration. In our situation, we must
be careful and cannot naively use a monomer or the center of gravity
of a polymer chain as a material point. In the deformed steady state,
chain diffusion processes carry on, and material points would
continue to move. Note, however, that in theoretical treatments
of viscoelasticity \cite{Doi86a,Bir77a}, one follows the motion of
monomers indeed, since when the pressure tensor is set up, one must
also consider the forces which are mediated by the polymer bonds.
A practicable alternative to characterize the deformed state is to
calculate collective quantities based on the monomer positions.
We proceed this way and divide the system into slices in the width
direction. In each slice, we compute the {\em center of the cross section},
\begin{equation}
  c (x) = \frac{1}{N(x)} \sum_{\rm \SY monomers \atop \rm \SY in \; slice}
          z_{\text{monomer}} \, ,
\end{equation}
as the average of the monomer $z$ coordinates in the slice at position $x$
($N(x)$ is the total number of monomers in that slice). The set of these
centers plotted over the width coordinate constitute what we call the
{\em bending line}. Furthermore, we consider the {\em radius of gyration}
of the cross section,
\begin{equation}
  r^2 (x) = \frac{1}{N(x)} \sum_{\rm \SY monomers \atop \rm \SY in \; slice}
          \left( z_{\text{monomer}} - c (x) \right)^2 \, .
\end{equation}
For films with approximately constant monomer density over the height
direction (and we take that for granted), this is a measure for the local
film thickness and a plot of $r(x)$ allows to detect regions of film
widening and strangling. Note that both definitions only use
the $z$ coordinates of the monomers, since $x$ represents the independent
coordinate, and the $y$ direction is neutral and used for averaging.
Note, however, that based on quantities which are solely averages over
single monomer coordinates (we may call this {\em statistics on points})
we cannot decide whether elastic bending or elastic or viscous shear
effects dominate the deformation of the polymer film. In order to proceed,
the {\em relative} positions of material points (which define lines or planes
in some reference configuration) must be considered. To discern elastic from
viscous effects, we must resolve the mapping of lines and planes by the
deformation process in time.

\subsection*{Bending lines}

A snapshot of one small configuration under bending is displayed
in Fig.~\ref{fig-02}. In order to show some layers of monomers
in the total front view, each monomer ball has been drawn with a
radius smaller than its Lennard--Jones radius of 3.8 {\AA}.
The section under the magnifying glass, however, uses this radius
for the monomers.
\begin{figure}
  \begin{center}
    \begin{minipage}{\textwidth}
      \insertps{0.6}{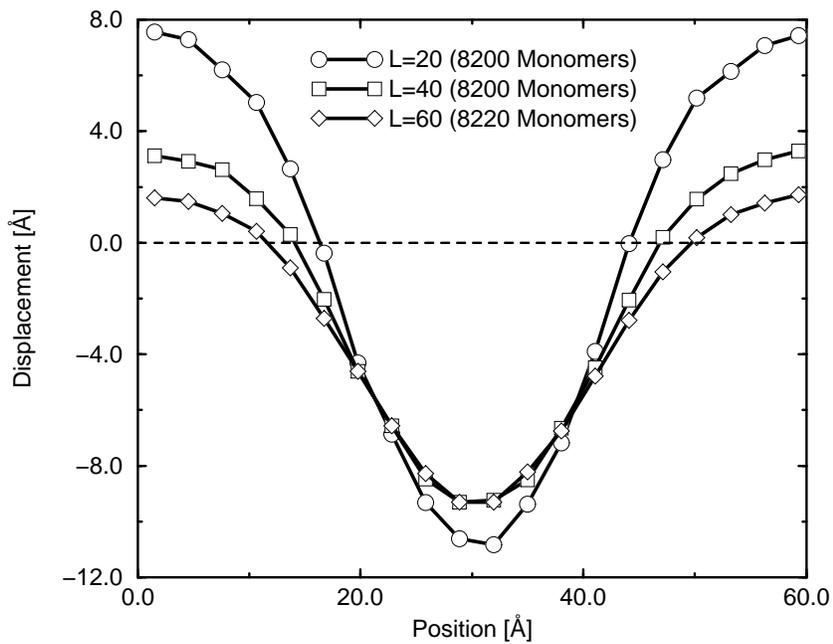}
    \end{minipage}
  \end{center}
  \caption{Bending lines for small systems.}
  \label{fig-04}
\end{figure}
The bending lines for small systems of various chainlengths are plotted
in Fig.~\ref{fig-04}. Note the different scales on both axes. For systems
of width 60 {\AA}, we find a maximum displacement of about 10 {\AA},
corresponding to a ratio of 6:1. We see that the system with the shortest
chains undergoes the largest deformation and that the material is not only
bent below the zero line but also displaced to both sides of the invading
tool. The difference between chainlength 20 and 40 is more prominent than
between 40 and 60.
This should be compared to the behaviour of medium systems, one of
which is shown in Fig.~\ref{fig-03}.
\begin{figure}
  \begin{center}
    \begin{minipage}{\textwidth}
      \insertps{0.6}{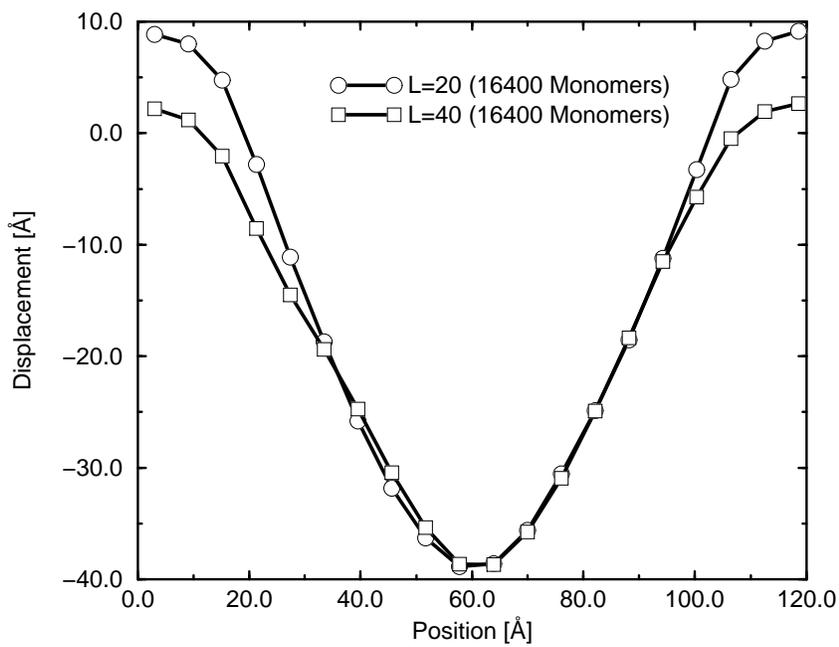}
    \end{minipage}
  \end{center}
  \caption{Bending lines for medium systems.}
  \label{fig-05}
\end{figure}
Their bending lines, Fig.~\ref{fig-05}, are more curved, the ratio
between width and maximum displacement is changed to 3:1. The difference
between the bending lines for chainlength 20 and 40 is reduced with
respect to the small configurations. Already from these obervations we
can infer that with the small configurations we merely perform indentation
experiments \cite{Hap96a}, at least for an applied force of 0.25 nN.
Due to the aspect ratio of the small systems, there is little space for
material displacement between the two support columns. On the other hand,
for the medium systems, a look at Fig.~\ref{fig-03} suggests that the
deformation is mainly limited by the constraint imposed by the lower plate
(which repells the polymer). We have already exceeded
the {\em terminal relaxation time} \cite{Gen79a}, and the polymer is
sheared like a liquid. For further simulations, we are urged to resolve
the initial, kinetic phase of the bending experiment more carefully if
we want to observe any elastic effects in accordance with the classical
notion of bending.

\subsection*{Film strangling}

\begin{figure}
  \begin{center}
    \begin{minipage}{\textwidth}
      \insertps{0.6}{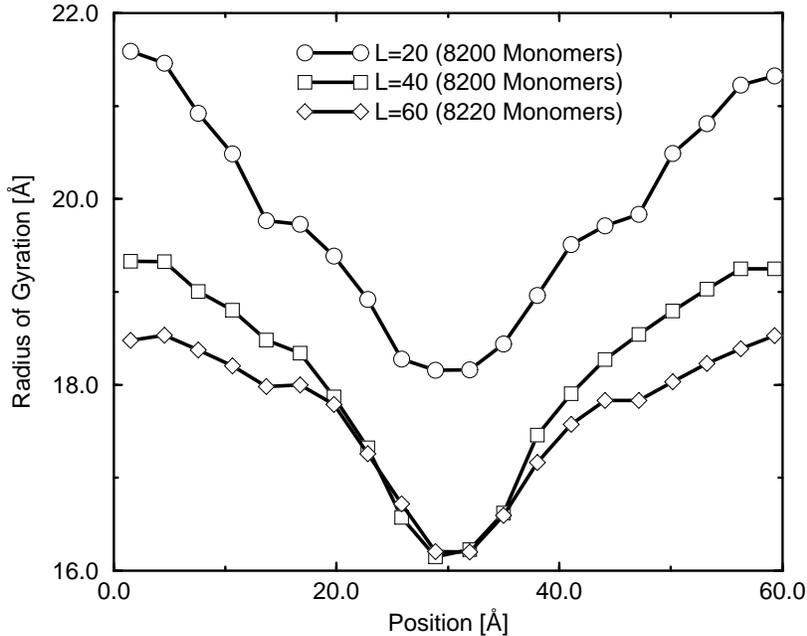}
    \end{minipage}
  \end{center}
  \caption{Radii of gyration for small systems.}
  \label{fig-06}
\end{figure}
The thickness of the polymer film locally changes under the action
of the invading tool. This can be concluded 
from plots of the radius of gyration over the width direction.
Fig.~\ref{fig-06} summarizes the situation for the small systems.
The equilibrium value for the radius in this plot is around 20 {\AA}.
We find film thickening at the sides and thinning in the center, below
the tool. This fits into our picture that these simulations correspond
to indentation experiments were material does not only move into the
cavity between the support columns but also gets displaced to both
sides of the tool. Again there is a significant difference between
chains of length 20 and of length 40. The longest chains show the
smallest displacement effects.
\begin{figure}
  \begin{center}
    \begin{minipage}{\textwidth}
      \insertps{0.6}{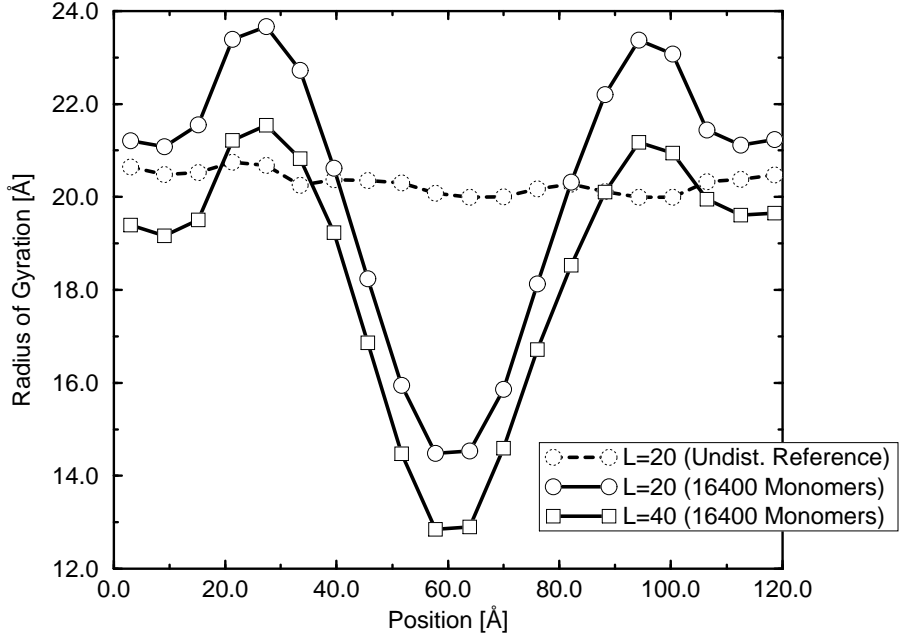}
    \end{minipage}
  \end{center}
  \caption{Radii of gyration for medium systems.}
  \label{fig-07}
\end{figure}
For medium systems we plot the strangling effects in Fig.~\ref{fig-07}.
They are more pronounced than for small ones. Also observe the clearly visible
thickening regions to both sides of the tool. Above the support columns,
the film becomes thinner again. This effect is plausible, since for systems
with larger width, the exact geometrical shape of the interacting body
(tool or support) ceases to matter and the interaction is reduced to an
overall displacement effect.

\subsection*{Indentation depth}

\begin{figure}
  \begin{center}
    \begin{minipage}{\textwidth}
      \insertps{0.6}{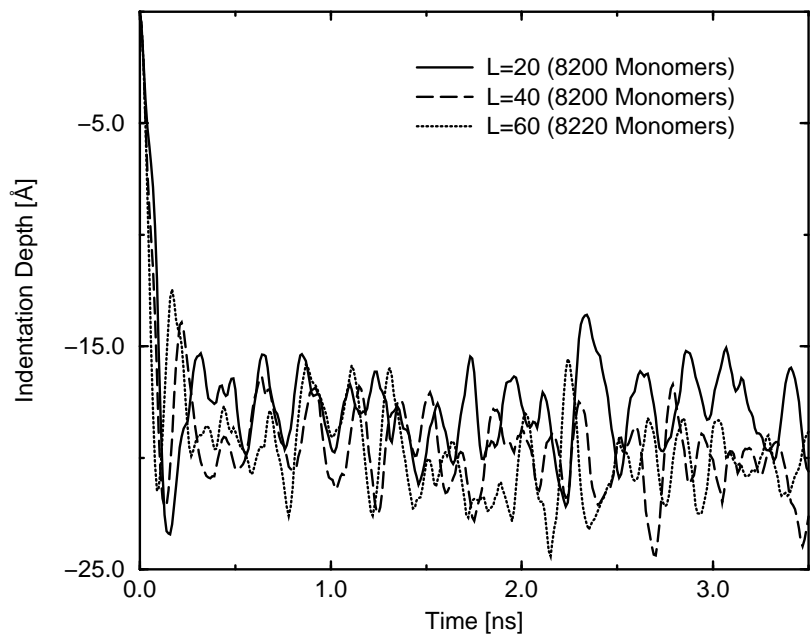}
    \end{minipage}
  \end{center}
  \caption{Time dependence of the tool indentation depth for small systems.}
  \label{fig-08}
\end{figure}

\begin{figure}
  \begin{center}
    \begin{minipage}{\textwidth}
      \insertps{0.6}{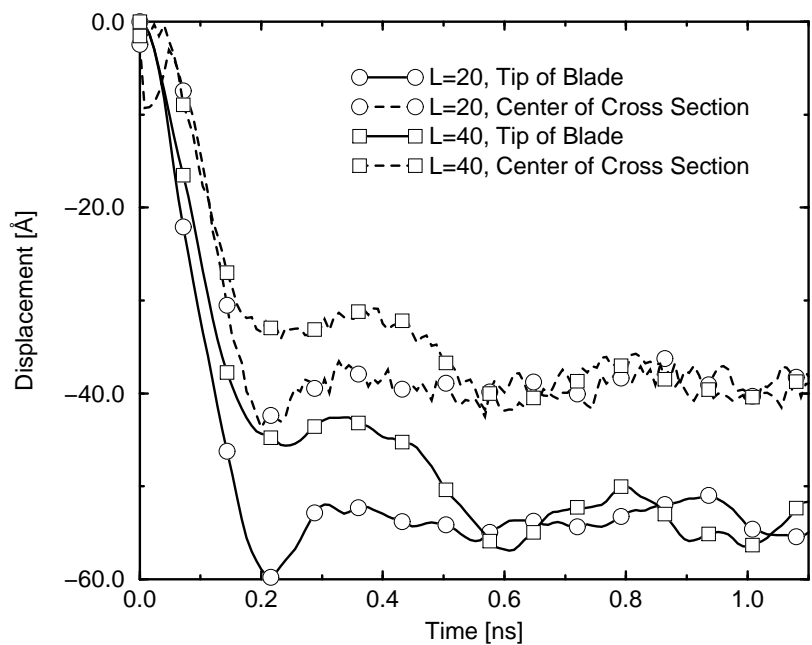}
    \end{minipage}
  \end{center}
  \caption{Connection between indentation depth and section displacement.}
  \label{fig-09}
\end{figure}

The indentation depth measures the movement of the tool's tip from
its starting position just above the polymer surface into the bulk.
The sign convention follows the orientation of the height axis and
the indentation counts negative if the tool moves down. 
In Fig.~\ref{fig-08}, we show the time development of this quantity
for three small systems with chainlength 20, 40, and 60. We conjecture
that the shorter the chains are, the lower lies the first minimum
of the curve (reflecting the higher flexibility of shorter chains)
but the higher is the average tip position in the steady state
(finally indicating a higher resistance against indentation of the
shorter chains). This is what also has been found in our dedicated
indentation simulations \cite{Hap96a}. Moreover, the position
fluctuations may well increase with decreasing chainlength.
To clarify the connection between the tip position and the deformation
of the polymer film, we have to think about some appropriate
norm of the displacement field. One such quantity certainly is
the maximum displacement found for some cross section (the minimum
of the bending curves). For medium configurations, this quantity together
with the tip position is plotted in Fig.~\ref{fig-09} for increasing time.
We see that this norm of the deformation state instantly follows the
external forcing, meaning that no delay effects which indicate viscoelasticity
can be observed on the time scale under consideration. Moreover, this figure
demonstrates that compared to small systems, the relative steady state
deformation fluctuations become reduced and that the difference between
chainlength 20 and 40 disappears in the steady state.

From Fig.~\ref{fig-08} we can estimate that the tool moves about 20 {\AA}
in 1 ns. This corresponds to an impact velocity of 2 m/s. Similarly, from
Fig.~\ref{fig-09} the estimate is 60 {\AA} in 0.2 ns, which means a tool
impact with 30 m/s or roughly 100 km/h. These time scales should be
compared to certain relaxation times which characterize equilibrium diffusion
as defined in Ref.~\cite{Pau91a}.  At the time $\tau_3$ in particular ,
determined by long term equilibrium runs and tabulated in Tab.~\ref{table-03},
the mean square displacement of mid chain monomers in the chains'
center of mass system is on average the same as the mean square displacement
of the center of mass itself (the former asymptotically approaches the
square of the radius of gyration whereas the latter increases linearly
in time). We see that especially for longer chainlengths there is no
strict separation of time scales (diffusion and indentation dynamics),
but the polymer has some time to adapt to the constraints imposed by
the tool. One is tempted to propose a szenario with three time stages:
the first stage includes the impact process, during the second
stage, the tool performs damped oscillations around some average
steady state position (which may, however, drift on some longer time
scale), and in the third stage, the polymer bulk restores equilibrium.
Note that in this paper we loosely refer to the ``steady state'' as to
the time when impact phenomena have faded out. 
\begin{table}
  \caption{Simulation values for the equilibrium relaxation time $\tau_3$.}
  \label{table-03}
  \begin{tabular}{ll}
    chainlength & time scale $\tau_3$ (see Ref.~\cite{Pau91a}) \\
  \tableline
    20 monomers & 8 ps \\
    40 monomers & 50 ps \\
    60 monomers & 240 ps
  \end{tabular}
\end{table}

\subsection*{Density profiles}

\begin{figure}
  \begin{center}
    \begin{minipage}{\textwidth}
      \insertps{0.6}{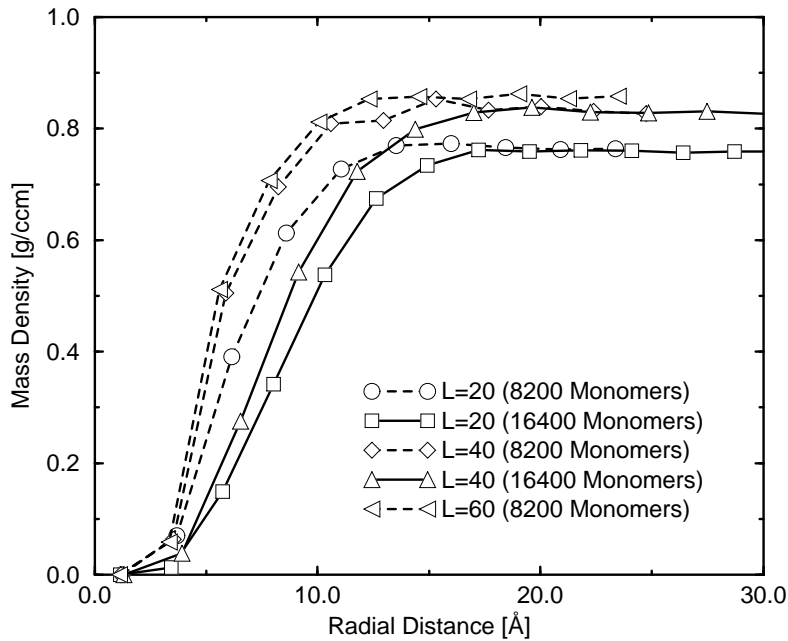}
    \end{minipage}
  \end{center}
  \caption{Radial mass density distribution.}
  \label{fig-10}
\end{figure}

If a solid body invades some yielding material, besides the gross
deformation also local effects are worth to investigate, especially
if one expects some transient phenomena or local defects
leading to failure (film fracture or strangling). As a first step towards
an accurate spatial resolution of the interaction zone between tool
and polymer, the polymer mass density found in shells of increasing
distance from the tool's tip is plotted in Fig.~\ref{fig-10}.
This plot combines small and medium systems. For all systems, the density
increases monotonically from zero (next to the tool) to its bulk value.
For the same chain length, both the small and the medium configurations
reach the same plateau value of the density which gives us additional
confidence that the systems are large enough to observe bulk behaviour.
Note that for small systems the diameter
of the shells is limited since otherwise parts of the shell lie outside
the region filled with material. For a given system size, we find that
the longer the chains, the higher the bulk density and the steeper the
flank of the density profile. For the medium systems, the profiles
are somewhat flatter, since in larger systems the material bends down
and has more room to move aside. Recall that the tool exerts a purely
repelling force with a length scale of $\sigma = \text{3.8 {\AA}}$ (which
is the same as the Lennard--Jones radius of the monomers).
The plateau density has been restored at a distance of roughly
$4 \sigma \approx \text{15 {\AA}}$ which should be compared to the
approximate film thickness of $20 \sigma \approx \text{75 {\AA}}$.

\section*{Conclusion and outlook}

We have demonstrated that the nano--mechanics of amorphous media like
polymer films can successfully be dealt with by computer simulation.
The length scale considered here is of particular interest, since we
treat a number of chains large enough to constitute some bulk, but, on
the other hand, too small to justify the naive application of continuum
mechanics. We have reported on illustrative but nevertheless
fundamental aspects of nano--deformation: displacement of cross sections,
film strangling, and indentation depth. These are the first steps towards
the kinematical characterization of the deformation process. We also took
a preliminary look at the local reaction of the material to the indentation
and displayed the mass density in the polymer bulk at increasing distance
from the tool. 

A step to be undertaken next is to resolve the non--equilibrium forces
which emerge during the deformation process, both in space and time.
It must clarified whether the notion of a local deformation field and
the concept of a stress can be suitably modified and adapted to the present
situation. We also must be explicit about the various dynamical processes
taking place in the computer experiment (diffusion, steady state deformation
fluctuations) and especially about their relaxation times {\em under
non--equilibrium conditions}. Of particular interest is the relation between
tool velocity and chain movement. A simplified picture of the deformation
process consisting of certain time stages (three of them have been suggested
above) can then be set up.
Finally, the long time behaviour of the polymer system under load is of
interest. One possible criterion to monitor are the non--equilibrium forces
in the polymer bulk caused by the invading tool. For long times, these forces
relax. It is highly probable however, that slow creep processes, controlled
by viscous steady state forces, carry on.
After the descriptive work is accomplished, one will be tempted to average
over the granularity of single chains and to set up a smoother model for
nano--deformation. An interesting prospect for theoretical work is then to
derive what corresponds to the constitutive equations of continuum mechanics
for such a model.

\section*{Acknowledgments}

This work is supported by the Bundesministerium f\"ur Bildung und
Forschung (BMBF) in the framework of the project ``Computer Simulation
Komplexer Materialien'' under grant no.~03N8008D. The authors have profited
a lot from fruitful discussions with members of the Bayer AG, Leverkusen.



\begin{references}
\bibitem{Hap95a} Th.~Hapke, {\it Simulation mikromechanischer Eigenschaften
  von amorphen makromolekularen Gl{\"a}sern am Beispiel von Polyethylen\/},
  Diplomarbeit am Institut f{\"u}r Theoretische Physik der Universit{\"a}t
  Heidelberg, 1995.
\bibitem{All87a} M.~P.~Allen and D.~J.~Tildesley, {\it Computer Simulation
  of Liquids\/} (Clarendon Press, Oxford, 1987).
\bibitem{Cha90a} A.~K~.Chakraborty, H.~T.~Davis, and M.~Tirrell,
    J.~Polym.~Sci.~A:~Polym.~Chem. {\bf 28}, 3185 (1990);
  J.~S.~Shaffer, A.~K~.Chakraborty, M.~Tirrell, H.~T.~Davis,
    and J.~L.~Martins, J.~Chem.~Phys.~{\bf 95}, 8616 (1991);
  T.~K.~Xia, J.~Ouyang, M.~W.~Ribarsky, and U.~Landman,
    Phys. Rev. Lett. {\bf 69}, 1967 (1992);
  W.~D.~Luedtke and U.~Landman, Comp. Mat. Sci. {\bf 1}, 1 (1992).
\bibitem{Doi86a} M.~Doi and S.~F.~Edwards, {\it The Theory of Polymer
  Dynamics\/} (Clarendon Press, Oxford, 1986).
\bibitem{Bir77a} R.~B.~Bird, R.~C.~Armstrong, O.~Hassager, and
  C.~F.~Curtiss, {\it Dynamics of Polymeric Liquids\/},
  vol.~1 (Fluid Mechanics) and vol.~2 (Kinetic Theory)
  (Wiley, New York, 1977).
\bibitem{Hap96a} Th.~Hapke, A.~Linke, and D.~W.~Heermann
  (to be published).
\bibitem{Gen79a} P.--G.~de~Gennes, {\it Scaling Concepts in Polymer
  Physics\/} (Cornell University Press, Ithaca, 1979).
\bibitem{Pau91a} W.~Paul, K.~Binder, D.~W.~Heermann, and K.~Kremer,
  J.~Chem.~Phys.~{\bf 95}, 7726 (1991).
\end{references}
\end{document}